\documentclass[twocolumn,english,prb]{revtex4}
\usepackage[T1]{fontenc}
\usepackage[latin9]{inputenc}
\usepackage{babel}
\usepackage{amsmath}
\usepackage{graphicx}
\usepackage{wasysym}
\usepackage[unicode=true,
 bookmarks=true,bookmarksnumbered=false,bookmarksopen=false,
 breaklinks=false,pdfborder={0 0 1},backref=false,colorlinks=false]
 {hyperref}
\hypersetup{pdftitle={Hierarchy of correlations for the Ising model in the Majorana representation},
 pdfauthor={Alvaro Gomez-Leon},
 pdfsubject={Physics}}
\usepackage{breakurl}

\makeatletter
\@ifundefined{textcolor}{}
{%
 \definecolor{BLACK}{gray}{0}
 \definecolor{WHITE}{gray}{1}
 \definecolor{RED}{rgb}{1,0,0}
 \definecolor{GREEN}{rgb}{0,1,0}
 \definecolor{BLUE}{rgb}{0,0,1}
 \definecolor{CYAN}{cmyk}{1,0,0,0}
 \definecolor{MAGENTA}{cmyk}{0,1,0,0}
 \definecolor{YELLOW}{cmyk}{0,0,1,0}
}

\makeatother

\begin{document}

\title{Hierarchy of correlations for the Ising model in the Majorana representation}

\author{Álvaro Gómez-León}

\affiliation{Department of Physics and Astronomy and Pacific Institute of Theoretical
Physics\\
University of British Columbia, 6224 Agricultural Rd., Vancouver,
B.C., V6T 1Z1, Canada.}

\date{\today}
\begin{abstract}
We study the quantum Ising model in D dimensions with the equation
of motion technique and the Majorana representation for spins. The
decoupling scheme used for the Green's functions is based on the hierarchy
of correlations in position space. To lowest order, this method reproduces
the well known mean field phase diagram and critical exponents. When
correlations between spins are included, we show how the appearance
of thermal fluctuations and magnons strongly affect the physical properties.
In 1D and for $B=0$ we demonstrate that, to first order in correlations,
thermal fluctuations completely destroy the ordered phase. For non
vanishing transverse field we show that the model exhibits different
behavior than its classical counterpart, specially near the quantum
critical point. We discuss the connection with the Dyson's equation
formalism and the explicit form of the self-energies.
\end{abstract}
\maketitle
\tableofcontents{}

\section*{Introduction}

The transverse Ising model corresponds to one of the most studied
systems in quantum physics. One reason is the large number of physical
effects that can be described within this model: Ferromagnetic/Paramagnetic
(FM/PM) transitions\cite{TransverseIsingBook}, spin glass phases\cite{Spinglass1,Spinglass2},
frustration\cite{Frustration1}, many-body localization\cite{Many-bodyLoc1},
etc. Besides its theoretical interest, it also models a large variety
of real systems, such as interacting magnetic molecules\cite{LiHoF1,LiHoF2},
coupled superconducting circuits and many others whose low energy
description consists on coupled two-level systems. In addition, the
understanding of the Ising system is very important for quantum computation;
for example, finding the ground state of the spin glass phase corresponds
to an NP problem\cite{IsingComplexity}, which is typically used to
benchmark quantum annealers\cite{D-wave2014,Google2016}. Furthermore,
the 2D Ising model in a transverse field corresponds to a universal
Hamiltonian\cite{UniversalIsing2D} which can be used to effectively
describe many other models. Related to all the previous interests,
recent the studies on the dynamical properties of the Ising model\cite{BastidasIsing,SpinBathAC},
or on the effect of decoherence\cite{Prokof'ev2000,DissipativeIsing2,DissipativeIsing1}
show how the physics is even richer than previously thought.

The equation of motion technique in combination with the hierarchy
of correlations represent a useful non-perturbative approach to understand
the consequences of correlations in many-body system\cite{Zubarev-GF60,Nagaoka-Kondo,FermionicGF-Spins,SCHREIBER-Disorder}.
When it is used to describe real space correlations, it corresponds
to the well known $1/\mathcal{Z}$ expansion, $\mathcal{Z}$ being
the coordination number of the lattice; however, it can also be generalized
to e.g., correlations between momentum space modes, providing an alternative
scaling which can improve the convergence under some circumstances\cite{Gomez-Leon2016}.
The main idea is to separate the correlation functions into uncorrelated
and correlated parts; here the first part describes the single particle
physics, while the second corresponds to collective excitation effects.
The usefulness of this separation becomes clear when one finds a scaling
for the correlated part, as a function of the physical parameters.
This allows to organize terms hierarchically, in terms of decreasing
contributions to the correlation functions.

In this work we analyze the transverse field Ising model in D dimensions,
using the hierarchy of correlations in position space. We discuss
the advantages and limitations of the method by comparing with some
well established results, and show that, as expected by the increase
of the coordination number with D, the scaling of real spatial correlations
improves with the dimension of the system. We also discuss the critical
exponents and the properties of the phase transitions in this model.

The work is organized as follows: In section I we introduce the Majorana
representation and the basic idea behind a hierarchy of correlations.
In section II we discuss the uncorrelated solution of the Quantum
Ising model and discuss its advantages and limitations. In section
III we include the effect of correlations in absence of a transverse
field, which allows for a simpler discussion of the method, and a
direct comparison with well known results for the classical Ising
model. In section IV we include the effect of correlations in the
Quantum Ising model and discuss its properties.

\section{Method}

The hierarchy of correlations can be easily implemented in terms of
double-time Green's functions, however, the particular exchange properties
of spins may lead to many different descriptions (one can use a mapping
to bosonic, fermionic or hardcore boson models). Here we consider
a description in terms of fermionic Green's functions to avoid the
$\omega=0$ pole problem\cite{FermionicGF-Spins}, in combination
with the Majorana representation for spins\cite{MajoranaRepresentation2,MajoranaRepresentation1}.
This allows to directly work with fermionic Green's functions and
operators, unifying the formalism and simplifying some of the calculations.
The Majorana representation of a spin $\vec{S}_{n}$ at site $n$
is given by:
\begin{equation}
S_{n}^{\alpha}=-\frac{i}{2}\epsilon_{\alpha\beta\gamma}\eta_{n}^{\beta}\eta_{n}^{\gamma}
\end{equation}
where $\eta_{n}^{\alpha}$ corresponds to a Majorana particle at site
$n$ with internal degree of freedom $\alpha=x,y,z$, and $\epsilon_{\alpha\beta\gamma}$
is the Levi-Civita symbol. The algebra of Majorana particles is characterized
by $\left\{ \eta_{n}^{\alpha},\eta_{m}^{\beta}\right\} =\delta_{\alpha,\beta}\delta_{n,m}$,
$\left(\eta_{n}^{\beta}\right)^{2}=\frac{1}{2}$, $\left(\eta_{n}^{\beta}\right)^{\dagger}=\eta_{n}^{\beta}$
and one advantage is that it reproduces the $SU\left(2\right)$ algebra,
preserving the rotational symmetry (which does not happen with e.g.,
the hard-core boson representation). As an example, the magnetization
along the z-axis simply corresponds to $\langle S_{n}^{z}\rangle=-i\langle\eta_{n}^{x}\eta_{n}^{y}\rangle$.
Note that another advantage of this representation is that it is local,
which makes simple the interpretation of the results in the original
language of spins.

To set up the hierarchy of correlations in position space one needs
to find the scaling properties of spatial correlations between spins.
For this we shall follow a previous derivation based on the monogamy
property of entanglement\cite{Gomez-Leon2016}. Consider the Ising
model in a transverse field Hamiltonian:
\begin{equation}
H=-B\sum_{i}S_{i}^{x}-\sum_{i,j>i}V_{i,j}S_{i}^{z}S_{j}^{z}
\end{equation}
where the first term corresponds to the coupling to a transverse magnetic
field $B$, and $V_{i,j}$ corresponds to the spin-spin interaction
(because the mapping to Majorana fermions is local, this discussion
can be done in the spin or Majorana fermion language). The large anisotropy
in the spin-spin interaction, is usually present in real experiments
due to the crystal field magnetization. Applying the Majorana representation
we obtain the next Hamiltonian:
\begin{equation}
H=iB\sum_{i}\eta_{i}^{y}\eta_{i}^{z}+\sum_{i,j>i}V_{i,j}\eta_{i}^{x}\eta_{i}^{y}\eta_{j}^{x}\eta_{j}^{y}
\end{equation}
For the study physical properties of the system we define the next
double-time Green's function:
\begin{equation}
G_{n,m}^{\alpha,\beta}\left(t,t^{\prime}\right)=-i\langle\eta_{n}^{\alpha}\left(t\right);\eta_{m}^{\beta}\left(t^{\prime}\right)\rangle
\end{equation}
where $\langle\ldots\rangle$ corresponds to the statistical average
with respect to a thermal density matrix $\hat{\rho}=e^{-\beta\hat{H}}$
at temperature $T=\beta^{-1}$, and $;$ indicates that we can use
either the time ordered, retarded or advanced Green's function, as
they all fulfill the same equation of motion. The temperature is usually
related with the Boltzmann distribution of a Markovian phonon bath
that is in equilibrium with the system via spin-phonon coupling, which
for simplicity we do not include in our description. We calculate
the Green's function from the Heisenberg equation of motion $i\partial_{t}\mathcal{O}=\left[H,\mathcal{O}\right]$,
which after a Fourier transformation to frequency domain reads:
\begin{eqnarray}
\omega G_{n,m}^{\alpha,\beta}\left(\omega\right) & = & \frac{\delta_{n,m}}{2\pi}\delta_{\alpha,\beta}+iB\epsilon_{x\alpha\theta}G_{n,m}^{\theta,\beta}\left(\omega\right)\nonumber \\
 &  & +\epsilon_{z\alpha\theta}\sum_{i\neq n}V_{n,i}G_{iin,m}^{xy\theta,\beta}\left(\omega\right)\label{eq:GeneralEOM}
\end{eqnarray}
where $G_{iin,m}^{xy\mu,\beta}\left(t,t^{\prime}\right)=-i\langle\eta_{i}^{x}\left(t\right)\eta_{i}^{y}\left(t\right)\eta_{n}^{\mu}\left(t\right);\eta_{m}^{\beta}\left(t^{\prime}\right)\rangle$.
Eq.\ref{eq:GeneralEOM} is the central object of this work, as the
Majorana double-time Green's function characterizes the order parameter
$\langle S_{n}^{z}\rangle$, and differentiates between the FM and
the PM phase. To find the scaling properties of correlations and obtain
a decoupling scheme for higher correlation functions let us consider
the diagonal part of the Green's function ($n=m$), which is related
with the on-site magnetization. The four-point function $G_{iin,n}^{xy\alpha,\beta}$,
proportional to the reduced density matrix $\rho_{i,n}$, can then
be separated into its uncorrelated and correlated parts according
to the general decomposition $\rho_{i,n}=\rho_{i}\rho_{n}+\rho_{i,n}^{C}$.
This allows to rewrite the equation of motion as:
\begin{eqnarray}
\omega G_{n,n}^{\alpha,\beta}\left(\omega\right) & = & \frac{\delta_{\alpha,\beta}}{2\pi}+iB\epsilon_{x\alpha\theta}G_{n,n}^{\theta,\beta}\left(\omega\right)\label{eq:CorrelaEq}\\
 &  & +\epsilon_{z\alpha\theta}\sum_{i\neq n}V_{n,i}\langle\eta_{i}^{x}\eta_{i}^{y}\rangle G_{n,n}^{\theta,\beta}\left(\omega\right)\nonumber \\
 &  & +\epsilon_{z\alpha\theta}\sum_{i\neq n}V_{n,i}\mathcal{G}_{iin,n}^{xy\theta,\beta}\left(\omega\right)\nonumber 
\end{eqnarray}
where $G_{n,n}^{\alpha,\beta}\propto\rho_{n}$ and $\mathcal{G}_{iin,n}^{\alpha\delta\gamma,\beta}\propto\rho_{n,i}^{C}$
correspond to the uncorrelated and correlated parts, respectively.
If the system is strongly correlated, the entanglement monogamy implies
that there is an upper bound to the amount of entanglement that a
set of spins can share, i.e. we cannot correlate the set with an extra
spin unless the entanglement in the initial set is reduced. In a system
with dominant nearest neighbors interaction, one would expect that
each spin is mainly entangled with its nearest neighbors; then if
we have to reduce the entanglement between them to include an extra
spin, it must decrease as $\rho_{i,n}^{C}\sim\mathcal{Z}^{-1}$; otherwise
we would violate the entanglement monogamy. Finally, it can be shown
that the scaling $\rho_{i,n}^{C}\sim\mathcal{Z}^{-1}$ implies $V_{i,j}\sim\mathcal{Z}^{-1}$
by calculating the equation of motion for the reduced density matrix.
The meaning of the $\mathcal{Z}^{-1}$ scaling is that contributions
due to entanglement between spins must scale inversely with the coordination
number. Finally, once we have fixed the scaling properties of correlations
$\rho_{i,n}^{C}\sim\mathcal{Z}^{-1}$ and $V_{i,j}\sim\mathcal{Z}^{-1}$,
we can organize the different terms in the equation of motion hierarchically
and find their solution. It is important to stress that the assumption
of a strongly correlated phase allows to derive an upper bound to
the scaling of correlations; however, if the phase is weakly correlated,
one would expect an even smaller contribution from the correlated
part.

To conclude this section, we discuss some general properties of the
quantum Ising model in equilibrium at $T=0$, which can be used to
check our results: (i) the transverse magnetization $\langle S_{i}^{y}\rangle=0$
to all orders of correlations, which can be proved from the calculation
of the equation of motion $\partial_{t}\langle S_{i}^{z}\rangle=0$;
(ii) the two-point correlation function $\langle S_{i}^{z}S_{j}^{y}\rangle=0$
to all orders, according to the equation of motion $\partial_{t}\langle S_{i}^{x}\rangle=0$.
Also, let us include the Fourier transform to $\mathbf{k}$-space
of Eq.\ref{eq:CorrelaEq}, which will be used in the last sections:
\begin{align}
\omega G_{\mathbf{k}}^{\alpha,\beta}\left(\omega\right) & =N\frac{\delta\left(\mathbf{k}\right)\delta_{\alpha,\beta}}{2\pi}+iB\epsilon_{x\alpha\theta}G_{\mathbf{k}}^{\theta,\beta}\left(\omega\right)\label{eq:k-space-GF}\\
 & +\epsilon_{z\alpha\theta}\frac{1}{N}\sum_{\mathbf{q}}V_{\mathbf{q}}\langle\eta_{\mathbf{q}}^{x}\eta_{\mathbf{q}}^{y}\rangle G_{\mathbf{k}-\mathbf{q}}^{\theta,\beta}\left(\omega\right)\nonumber \\
 & +\epsilon_{z\alpha\theta}\frac{1}{N}\sum_{\mathbf{q}}V_{\mathbf{q}}\mathcal{G}_{\mathbf{q},\mathbf{k}-\mathbf{q}}^{xy\theta,\beta}\left(\omega\right)\nonumber 
\end{align}
In the next section we will show that one can simplify the second
line, by assuming spatially homogeneous magnetization $\langle\eta_{\mathbf{q}}^{x}\eta_{\mathbf{q}}^{y}\rangle=\langle\eta^{x}\eta^{y}\rangle iN\delta\left(\mathbf{q}\right)=iM_{z}N\delta\left(\mathbf{q}\right)$
(this choice is discussed in the next section, but should hold for
ferromagnetic interaction and periodic boundary conditions).

\section{Uncorrelated solution}

\begin{figure}
\includegraphics[scale=0.8]{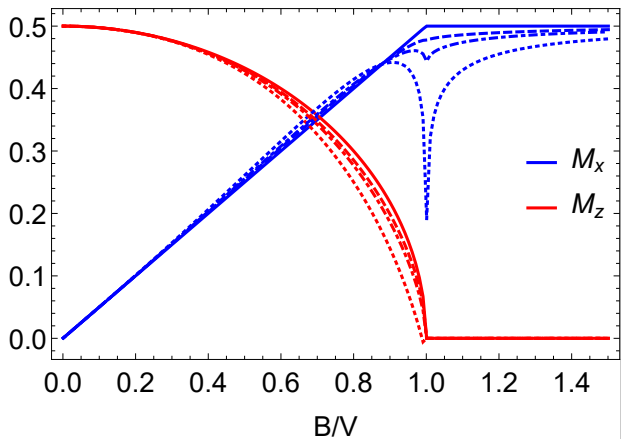}

\caption{\label{fig:PhaseDiagram1}The solid lines correspond to the uncorrelated
magnetization, as a function of the transverse field $B/V$ for $T=0$,
where we have re-scaled the field to $B\rightarrow B/D$ to match
the critical point in different dimensions. In dotted, dot-dashed
and dashed we plot the magnetization including correlations to lowest
order, as a function of external field, for $D=1,2$ and $3$, respectively.
The correlated solution in 1D shows large deviations from $\bar{M}_{x}$
around the critical point $B\apprle B_{c}$ due to large correlations,
however, they are reduced as $D$ increases. Importantly, the critical
point characterized by $M_{z}$ is unchanged by correlations, in agreement
with the exact solution of the transverse Ising model in 1D.}
\end{figure}

\begin{figure}
\includegraphics[scale=0.8]{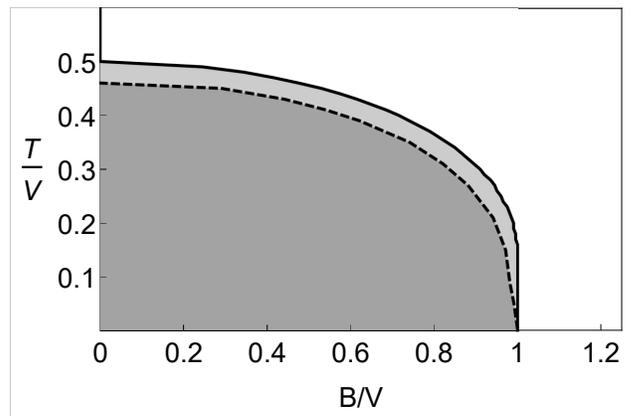}

\caption{\label{fig:PhaseDiagram2}The solid (dashed) line corresponds to the
uncorrelated (correlated to lowest order) phase diagram as a function
of the transverse field \textbf{$B/V$} and temperature $T/V$ for
a spin chain in arbitrary dimension (we re-scale $T\rightarrow T/D$
and $V\rightarrow V/D$ to make the critical lines coincide in absence
of correlations). Shaded area corresponds to the ferromagnetic phase.
We have omitted the phase diagram including correlations for the 2D
and the 3D case, as they are almost coincident with the mean field
result. This is in agreement with the decreasing role of correlations
in higher dimensions.}
\end{figure}

To characterize the ground state properties we are interested in the
magnetization $\langle S_{n}^{z}\rangle$, which is related with the
Majorana Green's function $G_{n,n}^{y,x}\left(t,t^{\prime}\right)$.
If we neglect all terms proportional to $\mathcal{Z}^{-1}$, we find
that the equation of motion, to lowest order, reduces to:
\begin{eqnarray}
\omega\bar{G}_{n,n}^{\alpha,\beta} & = & \frac{\delta_{\alpha,\beta}}{2\pi}+i\epsilon_{x\alpha\theta}B\bar{G}_{n,n}^{\theta,\beta}\label{eq:UncorrEq}\\
 &  & +\epsilon_{z\alpha\theta}\sum_{i\neq n}V_{n,i}\langle\eta_{i}^{x}\eta_{i}^{y}\rangle\bar{G}_{n,n}^{\theta,\beta}\nonumber 
\end{eqnarray}
where the bar on $\bar{G}_{n,n}^{\alpha,\beta}$ indicates the lowest
order contribution to the Green's function $G_{n,n}^{\alpha,\beta}$,
in powers of $1/\mathcal{Z}$. Note that the $\mathcal{Z}^{-1}$ scaling
coming from $V_{n,i}$ gets compensated by the sum over $\mathcal{Z}$
neighbors, and all terms in the equation are of order one. The solution
is easily found to be:
\begin{eqnarray}
\bar{G}_{n,n}^{y,x} & = & -\frac{\sum_{i\neq n}V_{n,i}\langle\eta_{i}^{x}\eta_{i}^{y}\rangle}{2\pi\left(\omega^{2}-\omega_{n}^{2}\right)}
\end{eqnarray}
where the poles of the Green's function are at $\pm\omega_{n}=\pm\sqrt{B^{2}-\left(\sum_{i\neq n}V_{n,i}\langle\eta_{i}^{x}\eta_{i}^{y}\rangle\right)^{2}}$.
Then the finite temperature, on-site magnetization, can be obtained
from the equal-time correlator\cite{Zubarev-GF60}:
\begin{eqnarray}
\langle\eta_{n}^{x}\eta_{n}^{y}\rangle & = & i\int\frac{\bar{G}_{n,n}^{y,x}\left(\omega+i\epsilon\right)-\bar{G}_{n,n}^{y,x}\left(\omega-i\epsilon\right)}{e^{\beta\omega}+1}d\omega\nonumber \\
 &  & =\frac{\sum_{i\neq n}V_{n,i}\langle\eta_{i}^{x}\eta_{i}^{y}\rangle}{2\omega_{n}}\tanh\left(\frac{\omega_{n}}{2T}\right)\label{eq:MF-Mag-Anisotropic}
\end{eqnarray}
Clearly this is a complicated self-consistency equation, as the magnetization
at site $n$ couples to the magnetization at the sites connected by
$V_{n,i}$. Hence, the solution clearly depends on the boundary condition
and on the details of the interaction. Assuming periodic boundary
conditions and ferromagnetic interaction, one can see that the solution
that minimizes the average energy $\langle H\rangle$ corresponds
to homogeneous magnetization. Hence, we drop the sub-index $n$ and
rewrite the equation in terms of the uncorrelated, average magnetization
$\bar{M}_{z}=-i\sum_{n}\langle\eta_{n}^{x}\eta_{n}^{y}\rangle/N$:
\begin{eqnarray}
\bar{M}_{z} & = & \frac{\bar{M}_{z}V_{0}}{2\omega_{0}}\tanh\left(\frac{\omega_{0}}{2T}\right)\label{eq:MF-Mz}
\end{eqnarray}
where $\omega_{0}=\sqrt{B^{2}+V_{0}^{2}\bar{M}_{z}^{2}}$ and $V_{\mathbf{k}}=\sum_{\mathbf{x}}e^{i\mathbf{k}\cdot\mathbf{x}}V_{\mathbf{x}}$.
Similarly, the calculation of the transverse magnetization results
in
\begin{eqnarray}
\bar{M}_{x} & = & \frac{B}{2\omega_{0}}\tanh\left(\frac{\omega_{0}}{2T}\right)\\
\bar{M}_{y} & = & 0
\end{eqnarray}
These equations reproduce the well know mean field self-consistency
equations of the Ising model\cite{Chakrabarti}, and the phase diagram
can be determined from their solution. We consider a 1D chain ($\mathcal{Z}=2$),
a square lattice ($\mathcal{Z}=4$) and a cubic lattice ($\mathcal{Z}=6$),
and plot in Fig.\ref{fig:PhaseDiagram1}(solid line) the value of
the magnetization as a function of the transverse field at $T=0$.
It shows that, for vanishing transverse field, all spins are aligned
in parallel, minimizing the mean-field energy per spin $\langle H\rangle/N=-\bar{M}_{z}^{2}V_{0}$.
As the transverse field increases, quantum fluctuations $|\uparrow\rangle\leftrightarrow|\downarrow\rangle$
are produced, and for $B>V_{0}/2$ the longitudinal magnetization
vanishes. This is the hallmark of the paramagnetic/ferromagnetic transition.
Fig.\ref{fig:PhaseDiagram2}(solid line) corresponds to the finite-temperature
phase diagram for the FM/PM phase transition, where now the temperature
produces thermal fluctuations, destroying the ferromagnetic phase
at the Curie temperature $T_{c}$.
From the uncorrelated results one finds the critical points expected
from the mean field solution of the Ising model: $B_{c}\left(T=0\right)=V_{0}/2$
and $T_{c}\left(B=0\right)=V_{0}/4$, where $\mathcal{Z}V=V_{0}$.
Note that in absence of correlations, the lattice dimension does not
affect the result (just $\mathcal{Z}$ is present), which usually
leads to violations of the no-go theorems for quantum phase transitions
in 1D and 2D\cite{Mermin-Wagner,1DPhaseTransitions}. For example,
this happens in the 1D classical Ising model ($B=0$ in the present
Hamiltonian), where it is known that the ferromagnetic phase is destroyed
by thermal fluctuations. We will see in the next section that the
addition of interspin correlations in the equation of motion allows
to capture this feature. For the 1D Quantum Ising model one can exactly
solve the equations using the Wigner-Jordan transformation, and this
allows us to compare our results with the exact solution. In this
case, the $T=0$ critical point is already reproduced by the uncorrelated
solution, although the quasi-particle spectrum and critical properties
are not\cite{IsingPhaseDiag}.

It is worth to stress that we have highly simplified the mean field
solution by assuming ferromagnetic interaction and periodic boundary
conditions in Eq.\ref{eq:MF-Mag-Anisotropic}. In general, the mean
field solution can be very powerful and it would capture interesting
effects, as for example the dimerization induced by anti-ferromagnetic
interaction, or frustration for some specific boundary conditions.

\paragraph*{Critical properties in absence of correlations:}

It is known that near a quantum critical point, the system's properties
can be characterized solely by its critical exponents \cite{Hertz-Universality}
and that this feature leads to the important concept of universality
classes. Here we discuss some of the critical properties of the hierarchy
in absence of correlations. Although all mean field theories are characterized
by the same critical exponents, independently of the system's dimension,
we will first show how one can obtain their value from the previous
results, and then in the next sections, compare with the case when
interspin correlations are included.

In absence of correlations, we can use the self-consistency equation
(Eq.\ref{eq:MF-Mz}) to calculate the critical exponent $\beta$ near
$T_{c}$ and $B_{c}$: In the FM phase and near the critical temperature
we have $t,\bar{M}_{z}\ll1$, where $t=\left(T_{c}-T\right)/T_{c}$
is the reduced temperature. Then the magnetization can be approximated
as ($B=0$):
\begin{equation}
\bar{M}_{z}=\frac{1}{2}\tanh\left(\frac{2\bar{M}_{z}}{1-t}\right)\simeq\frac{\bar{M}_{z}}{1-t}-\frac{1}{6}\left(\frac{2\bar{M}_{z}}{1-t}\right)^{3}+\mathcal{O}\left(\bar{M}_{z}^{5}\right)
\end{equation}
where we have used the series expansion $\tanh\left(x\right)\simeq x-\frac{x^{3}}{3}+\mathcal{O}\left(x^{5}\right)$.
The solutions to lowest order in $t$ are:
\begin{equation}
\bar{M}_{z}=0,\ \bar{M}_{z}=\pm\frac{\sqrt{3t}}{2}
\end{equation}
which imply that the critical exponent for the order parameter near
$T_{c}$ is $\beta_{t}=1/2$. If in a similar manner we analyze the
critical behavior near $B_{c}$, as a function of the reduced field
$b=\left(B_{c}-B\right)/B_{c}$. We find:
\begin{equation}
\bar{M}_{z}^{2}=\frac{b}{2}-\frac{b^{2}}{4}\simeq\frac{b}{2}
\end{equation}
and the solutions for the magnetization are:
\begin{equation}
\bar{M}_{z}=0,\ \bar{M}_{z}=\pm\sqrt{\frac{b}{2}}
\end{equation}
with the corresponding critical exponent $\beta_{b}=1/2$. Note that
we differentiate the critical exponent near $B_{c}$ and $T_{c}$
because the first corresponds to a quantum critical point, while the
second to a classical one. As expected for mean field solutions, both
relevant parameters have identical critical exponents $\beta=1/2$.
The anomalous dimension $\eta$ and the correlation length $\xi$
must be obtained from the correlated part of the two-point function
$\langle S_{\mathbf{k}}^{z}S_{\mathbf{k}^{\prime}}^{z}\rangle_{C}$,
which vanishes at this order of the hierarchy \textendash{} This is
in contrast with some refined calculations of the mean field solution,
where the fluctuation-dissipation theorem is used, giving a non-vanishing
correlated part\cite{Bellac}. Equivalently, corresponds to the difference
between Landau and Ginzburg\textendash Landau theory. We will show
in the next section that this is captured adding first order perturbations
to the uncorrelated solution.

\section{Correlations in absence of transverse field}

Now we consider the next order in the hierarchy of correlations, and
include $\mathcal{O}\left(\mathcal{Z}^{-1}\right)$ terms. We expect
that the corrections due to correlations will become more important
as we approach the phase boundary, as it is known that at the phase
boundary, the system becomes critical and interspin correlations diverge.
Furthermore, as $\mathcal{Z}$ changes with the dimension, we would
expect that the scaling of corrections due to correlations is worse
in the 1D case, while 2D and 3D should display smaller contributions.

Now the term $\mathcal{G}_{iin,n}^{xy\mu,\beta}$ that was neglected
in absence of correlations in Eq.\ref{eq:CorrelaEq} contributes and
must be included. We calculate the equation of motion for the correlated
part of the four-point function $G_{ppn,n}^{xy\alpha,\beta}=-i\langle\eta_{p}^{x}\eta_{p}^{y}\eta_{n}^{\alpha};\eta_{n}^{\beta}\rangle$
in the Appendix. This is done by calculating the equation of motion
for the four-point function, removing the uncorrelated part $\langle\eta_{p}^{x}\eta_{p}^{y}\rangle G_{n,n}^{\alpha,\beta}$,
and neglecting terms of $\mathcal{O}\left(\mathcal{Z}^{-2}\right)$
or higher. The solution for the four-point function, including the
transverse field, can be obtained analytically, however it is too
long to be explicitly written here. For simplicity let us first consider
the model in absence of a transverse field ($B=0$). In this case
the solution for the correlated part highly simplifies, and one finds
that the Fourier transform of $\mathcal{G}_{iin,n}^{xy\mu,\beta}$
is:
\begin{equation}
\mathcal{G}_{\mathbf{k},\mathbf{k}^{\prime}}^{xy\alpha,\beta}=-\frac{\omega\Lambda_{\mathbf{k},\mathbf{k}^{\prime}}^{\alpha,\beta}+\epsilon_{z\alpha\theta}iV_{0}M_{z}\Lambda_{\mathbf{k},\mathbf{k}^{\prime}}^{\theta,\beta}}{\omega^{2}-V_{0}^{2}M_{z}^{2}}\label{eq:GCorr-Classical}
\end{equation}
where
\begin{eqnarray}
\Lambda_{\mathbf{k},\mathbf{k}^{\prime}}^{\alpha,\beta} & = & \epsilon_{z\alpha\theta}\left(\frac{1}{4}-M_{z}^{2}\right)V_{-\mathbf{k}}G_{\mathbf{k}+\mathbf{k}^{\prime}}^{\theta,\beta}\\
 &  & -\frac{\epsilon_{z\alpha\theta}}{N}\sum_{\mathbf{q}}V_{\mathbf{q}}\langle\eta_{\mathbf{k}}^{x}\eta_{\mathbf{k}}^{y}\eta_{\mathbf{q}}^{x}\eta_{\mathbf{q}}^{y}\rangle_{C}G_{\mathbf{k}^{\prime}-\mathbf{q}}^{\theta,\beta}\nonumber 
\end{eqnarray}
At this point, the simplest approximation that one can make is to
substitute the Green's functions $G_{\mathbf{k}}^{\theta,\beta}$
in $\Lambda_{\mathbf{k},\mathbf{k}^{\prime}}^{\alpha,\beta}$, by
its uncorrelated solution $\bar{G}_{\mathbf{k}}^{\theta,\beta}$.
Then one can solve the self-consistency equation for the equal-time
correlator $\langle\eta_{\mathbf{k}}^{x}\eta_{\mathbf{k}}^{y}\eta_{\mathbf{k}^{\prime}}^{x}\eta_{\mathbf{k}^{\prime}}^{y}\rangle_{C}$
and find:
\begin{align}
\langle S_{\mathbf{k}}^{z}S_{\mathbf{k}^{\prime}}^{z}\rangle_{C}= & -\langle\eta_{\mathbf{k}}^{x}\eta_{\mathbf{k}}^{y}\eta_{\mathbf{k}^{\prime}}^{x}\eta_{\mathbf{k}^{\prime}}^{y}\rangle_{C}\label{eq:ZZ-Correlation1}\\
= & \frac{N\delta_{\mathbf{k},-\mathbf{k}^{\prime}}V_{\mathbf{k}^{\prime}}\left(\frac{1}{4}-\bar{M}_{z}^{2}\right)}{4T\cosh^{2}\left(\frac{V_{0}\bar{M}_{z}}{2T}\right)-V_{\mathbf{k}^{\prime}}}\nonumber 
\end{align}
As expected in absence of the transverse field (i.e., quantum fluctuations),
the correlation function $\langle S_{\mathbf{k}}^{z}S_{\mathbf{k}^{\prime}}^{x}\rangle_{C}$
vanishes. On the other hand, longitudinal correlations $\langle S_{\mathbf{k}}^{z}S_{\mathbf{k}^{\prime}}^{z}\rangle_{C}$
vanish at $T=0$, where the exact $\mathbf{Z}_{2}$ ground state of
the system is given by a product state of polarized spins along $z$,
as well as for $T\rightarrow\infty$, where the ground state corresponds
to an incoherent set of decoupled spins. The transition mechanism
between the two limits corresponds to a competition between thermal
fluctuations and spin-spin interactions. It is interesting the appearance
of a mass term in Eq.\ref{eq:ZZ-Correlation1} due to the solution
of the self-consistency equation, which is proportional to $T-T_{c}$
(just by expanding $V_{\mathbf{k}}$ in powers of $\mathbf{k}$ and
setting $V_{0}=4T_{c}$); this is very similar to the mass term obtained
in $\phi^{4}$-theory when the Hubbard-Stratonovich transformation
is applied to the Ising model.

\begin{figure}
\includegraphics[scale=0.8]{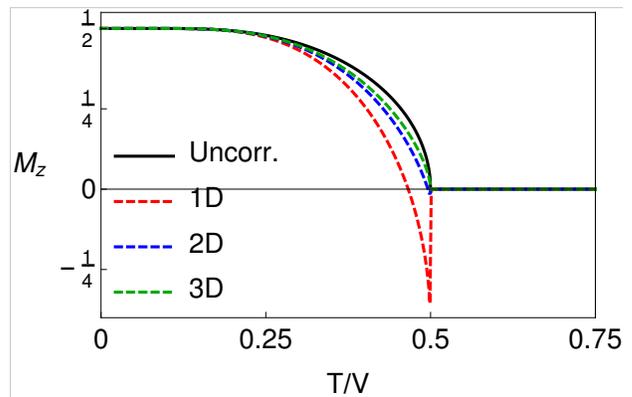}

\caption{\label{fig:Average-magnetization-Correlated-1}Numerical calculation
of the average magnetization including lowest order correlations,
as a function of $T$ for $D=1,2,3$ (red,blue and green, respectively).
In $D=1$ we obtain an unphysical remnant magnetization, which is
cured by including feedback (Fig.\ref{fig:SCE-Full}). For $D\geq2$
the discontinuity is washed out and the corrections smaller. The critical
point has been re-scaled to $T_{c}\rightarrow T_{c}/D$, to make it
coincident for all cases.}
\end{figure}

From Eq.\ref{eq:k-space-GF} it is clear that the magnetization will
now include corrections due to interspin correlations. Inserting Eq.\ref{eq:ZZ-Correlation1}
into Eq.\ref{eq:k-space-GF} we find:
\begin{equation}
M_{z}=\bar{M}_{z}-\left(\frac{1}{4}-\bar{M}_{z}^{2}\right)\frac{1}{N}\sum_{\mathbf{q}}\frac{\frac{\left|V_{\mathbf{q}}\right|^{2}}{2T}\tanh\left(\frac{V_{0}\bar{M}_{z}}{2T}\right)}{4T\cosh^{2}\left(\frac{V_{0}\bar{M}_{z}}{2T}\right)-V_{\mathbf{q}}}\label{eq:Classical-Corr-Mag}
\end{equation}
where $\bar{M}_{z}$ is the average magnetization, previously obtained
in absence of correlations. The sign of the correction term shows
that interspin correlations mostly act against the formation of ferromagnetic
order, and as in 1D these corrections will be larger, we expect that
the FM phase should be significantly reduced. Its numerical calculation
for different dimensions is shown in Fig.\ref{fig:Average-magnetization-Correlated-1}
(the divergence due to a vanishing denominator in Eq.\ref{eq:Classical-Corr-Mag}
is automatically cured by the input of the mean field values $\bar{M}_{z}$).
It shows that due to thermal fluctuations, the ferromagnetic phase
shrinks down for all cases, and in 1D it changes sign close to $T_{c}$.
This result is unphysical and it is produced due to the poor scaling
of correlations in 1D, where $\mathcal{Z}=2$. As we approach the
critical point, correlations between spins increase, and contributions
from $\mathcal{Z}^{-n}$ terms, with $n>2$, become important. The
fact that in 1D we observe a stronger influence of correlations is
also related with the fact that the ferromagnetic phase in 1D should
be destroyed by the thermal fluctuations at arbitrary $T\neq0$ and
the two main sources of this failure are: (1) The use of the uncorrelated
Green's functions in Eq.\ref{eq:GCorr-Classical}, neglecting the
feedback of the magnetization in the interspin correlations, and (2)
higher n-point functions will become increasingly dominant according
to renormalization group arguments. The first issue can be easily
fixed by rewriting the equation of motion as a Dyson's equation, as
we show next. The implementation of a renormalization group analysis
will be performed in future publications.

To include back-reaction between the renormalized magnetization and
the interspin correlations we insert Eq.\ref{eq:GCorr-Classical}
into Eq.\ref{eq:k-space-GF}. As the magnetization is spatially homogeneous
the equation of motion can be written as:
\begin{eqnarray}
\omega G_{0}^{\alpha,\beta} & = & \frac{\delta_{\alpha,\beta}}{2\pi}+\epsilon_{z\alpha\theta}iM_{z}V_{0}G_{0}^{\theta,\beta}\\
 &  & +\chi\frac{\omega G_{0}^{\alpha,\beta}+\epsilon_{z\alpha\theta}iV_{0}M_{z}G_{0}^{\theta,\beta}}{\omega^{2}-V_{0}^{2}M_{z}^{2}}\nonumber 
\end{eqnarray}
where we have defined
\begin{equation}
\chi=\frac{1}{N}\sum_{\mathbf{q}}\left|V_{\mathbf{q}}\right|^{2}\left[\frac{1}{4}-M_{z}^{2}-\frac{1}{N}\langle\eta_{\mathbf{q}}^{x}\eta_{\mathbf{q}}^{y}\eta_{-\mathbf{q}}^{x}\eta_{-\mathbf{q}}^{y}\rangle_{C}\right]
\end{equation}
and $G_{0}^{\alpha,\beta}=G_{\mathbf{k}=0}^{\alpha,\beta}/N$. This
can be written in matrix form as follows:
\begin{equation}
\omega\hat{G}=\frac{\hat{\delta}}{2\pi}+\hat{H}_{0}\cdot\hat{G}+\hat{\Sigma}\left(\omega\right)\cdot\hat{G}\label{eq:DysonEq1}
\end{equation}
and if we define the uncorrelated Green's function matrix as $\hat{G}^{\left(0\right)}=\left[\omega-\hat{H}_{0}\right]^{-1}$,
we find the usual Dyson's like equation:
\begin{equation}
\hat{G}=\hat{G}^{\left(0\right)}\cdot\frac{\hat{\delta}}{2\pi}+\hat{G}^{\left(0\right)}\cdot\hat{\Sigma}\left(\omega\right)\cdot\hat{G}
\end{equation}
with frequency dependent self-energy
\begin{equation}
\hat{\Sigma}\left(\omega\right)=\chi\frac{\hat{\sigma}_{0}\omega-\hat{\sigma}_{y}V_{0}M_{z}}{\omega^{2}-V_{0}^{2}M_{z}^{2}}\label{eq:Class-Self-energy}
\end{equation}
being $\hat{\delta}=\left(\delta_{x,\beta},\delta_{y,\beta}\right)^{T}$,
$\sigma_{y}=\left(\begin{array}{cc}
0 & -i\\
i & 0
\end{array}\right)$ and $\hat{\sigma}_{0}$ the identity matrix. The solution for the
Green's function can be obtained by direct matrix inversion:
\begin{equation}
\hat{G}=\left[\omega-\hat{H}_{0}-\hat{\Sigma}\left(\omega\right)\right]^{-1}\cdot\frac{\hat{\delta}}{2\pi}\label{eq:FullGF-Classical}
\end{equation}
We can now compare the full solution (Eq.\ref{eq:FullGF-Classical})
with the one obtained in Eq.\ref{eq:Classical-Corr-Mag}. It is equivalent
to the substitution of $\hat{G}$ by $\hat{G}^{\left(0\right)}$ on
the rhs of Eq.\ref{eq:DysonEq1}, and then equivalent to first order
perturbation over the mean field value. This explains the breakdown
for the magnetization as one approaches the phase boundary in Fig.\ref{fig:Average-magnetization-Correlated-1}.

In Eq.\ref{eq:FullGF-Classical} back-reaction between $G$ and $\mathcal{G}$
is now included, and although the formal solution seems simple, the
calculation of the self-energy can be quite complicated. The reason
is that it contains terms proportional to $\langle\eta_{\mathbf{q}}^{x}\eta_{\mathbf{q}}^{y}\eta_{-\mathbf{q}}^{x}\eta_{-\mathbf{q}}^{y}\rangle_{C}$,
which must be determined self-consistently with $M_{z}$. As the purpose
of this manuscript is to discuss the application of the hierarchy
of correlations to general spins systems, and not to fully characterize
the Ising model (for which a large number of accurate results is already
available), we will only analyze the full solution for the 1D case,
proving that the suppression of the ferromagnetic phase is captured
in very simple terms. Results in higher dimensions should be even
more accurate, which shows the power of this approach.

As we will discuss below for the general case, one of the advantages
of the Ising model with $B=0$ is that the frequency dependence of
the self-energy can be pulled out of the integral over $\mathbf{q}$.
This allows to exactly calculate the poles of the Green's function
when interspin correlations are included. In the presence of correlations,
the poles obtained in absence of correlations separate in pairs, and
their splitting is proportional to the strength of correlations $\omega=\pm V_{0}M_{z}\pm\sqrt{\chi}$
. We expect that the addition of higher spin correlations will induce
new poles as well, approaching a continuum in the real axis, as it
is expected in the thermodynamic limit.

Before we fully solve the self-consistency equation, let us consider
a slightly easier calculation, which also provides some insight into
the effect of correlations in the Ising model. This consists in approximating
the statistical average in the self-energy $\langle\eta_{\mathbf{q}}^{x}\eta_{\mathbf{q}}^{y}\eta_{-\mathbf{q}}^{x}\eta_{-\mathbf{q}}^{y}\rangle_{C}$
by its lowest order expression (Eq.\ref{eq:ZZ-Correlation1}):
\begin{equation}
\chi=\left(\frac{1}{4}-M_{z}^{2}\right)\frac{1}{N}\sum_{\mathbf{q}}\frac{4T\cosh^{2}\left(\frac{V_{0}M_{z}}{2T}\right)\left|V_{\mathbf{q}}\right|^{2}}{4T\cosh^{2}\left(\frac{V_{0}M_{z}}{2T}\right)-V_{-\mathbf{q}}}\label{eq:LowestOrder-SE}
\end{equation}
In this case one neglects the self-consistency equation for $\langle\eta_{\mathbf{q}}^{x}\eta_{\mathbf{q}}^{y}\eta_{-\mathbf{q}}^{x}\eta_{-\mathbf{q}}^{y}\rangle_{C}$
and just $M_{z}$ is determined self-consistently. This solution corresponds
to what one would expect from the Random Phase Approximation (RPA).
The explicit calculation of Eq.\ref{eq:LowestOrder-SE} diverges when
$4T\cosh^{2}\left(\frac{V_{0}M_{z}}{2T}\right)<V_{0}$; this is an
instability inherent to the RPA calculation, indicating that the ground
state is unstable with respect to these collective excitations\cite{MBExposed}
(because thermal excitations will shift the phase boundary or even
destroy the ferromagnetic phase). Generally the divergence is corrected
by the addition of the collective mode contributions to the self-energy,
and in our case, it will be cured by including the self-consistency
equation for $\langle\eta_{\mathbf{q}}^{x}\eta_{\mathbf{q}}^{y}\eta_{-\mathbf{q}}^{x}\eta_{-\mathbf{q}}^{y}\rangle_{C}$,
instead of using its lowest order approximation.

\begin{figure}
\includegraphics[scale=0.8]{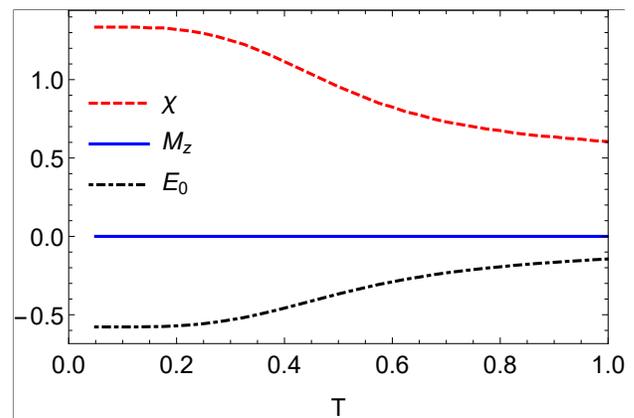}

\caption{\label{fig:SCE-Full}Full solutions to the self-consistency equations
and average energy as a function of $T/V$ for the 1D classical Ising
model.}
\end{figure}

Now to obtain the full solution we calculate $M_{z}$ from the Green's
function in Eq.\ref{eq:FullGF-Classical}, and $\chi$ from Eq.\ref{eq:GCorr-Classical}
with $G_{0}^{\alpha,\beta}$ from Eq.\ref{eq:FullGF-Classical}. After
some simple manipulations one finds:
\begin{eqnarray}
M_{z} & = & \frac{\frac{1}{2}\sinh\left(\frac{V_{0}M_{z}}{T}\right)}{\cosh\left(\frac{V_{0}M_{z}}{T}\right)+\cosh\left(\frac{\sqrt{\chi}}{T}\right)}\label{eq:SCE-MZ-Full}\\
\chi & = & \left(\frac{1}{4}-M_{z}^{2}\right)\frac{1}{N}\sum_{\mathbf{q}}\frac{m\left|V_{\mathbf{q}}\right|^{2}}{m-\frac{V_{\mathbf{q}}}{V_{0}}}\label{eq:SCE-CHI-Full}
\end{eqnarray}
with $m=2\sqrt{\chi}\left[\cosh\left(\frac{M_{z}V_{0}}{T}\right)+\cosh\left(\frac{\sqrt{\chi}}{T}\right)\right]/V_{0}\sinh\left(\frac{\sqrt{\chi}}{T}\right)$.
Eqs.\ref{eq:SCE-MZ-Full} and \ref{eq:SCE-CHI-Full} are one of the
main results. They correspond to the generalization of the mean field
equation for the magnetization, now including the effect of interspin
correlations. They are valid in arbitrary dimension, and both reduce
to the mean field case if $\chi\rightarrow0$. It is important to
note that Eq.\ref{eq:SCE-CHI-Full} has a divergence, however its
different to the RPA one previously obtained, and basically restricts
$\chi$ to positive values. This is physically meaningful because
$\chi$ corresponds to the integral of $\langle S_{\mathbf{k}}^{z}S_{-\mathbf{k}}^{z}\rangle_{C}$
over the whole Brillouin zone, which is always positive because they
tend to be aligned. The positivity of $\chi$ also makes the poles
of the Green's function to be real valued. Finally, we also include
the calculation of the average energy per spin (details of the calculation
in the Appendix):
\begin{equation}
E_{0}=-V_{0}M_{z}^{2}-\frac{\frac{1}{2}\sqrt{\chi}\sinh\left(\frac{\sqrt{\chi}}{T}\right)}{\cosh\left(\frac{V_{0}M_{z}}{T}\right)+\cosh\left(\frac{\sqrt{\chi}}{T}\right)}\label{eq:Energy-Class-Ising}
\end{equation}
The second term shows that interspin correlations always reduce the
energy and stabilize the system. Now we particularize the previous
expressions to 1D and study the fate of the ferromagnetic phase when
correlations are included. The self-consistency equation in the continuum
limit for $\chi$ (Eq.\ref{eq:SCE-CHI-Full}) can be calculated analytically
by contour integration in the complex plane, leading to ($V_{\mathbf{q}}=V_{0}\cos\left(q\right)$):
\begin{equation}
\chi_{1D}=\frac{\left(\frac{1}{4}-M_{z}^{2}\right)m^{2}V_{0}^{2}}{m^{2}-1+m\sqrt{m^{2}-1}}
\end{equation}
Fig.\ref{fig:SCE-Full} shows the solution to the self-consistency
equations Eq.\ref{eq:SCE-MZ-Full} and Eq.\ref{eq:SCE-CHI-Full}.
We find that the phase with $M_{z}=0$ is the one that persists in
the presence of correlations, to arbitrary low temperature. We also
plot the ground state energy as a function of temperature, and it
shows that as the temperature decreases, interspin correlations increase
and the energy of the paramagnetic ground state is reduced. Interestingly
the self-consistency equations also have a solution corresponding
to a first order transition for $T\sim0.25V$, but its energy is higher
than the one for $M_{z}=0$.

These solutions to the self-consistency equations fully characterize
the Green's functions in 1D. Higher dimensional cases can be solved
in a similar way, with the difference that one does not have an analytical
expression for the integral in Eq.\ref{eq:SCE-CHI-Full} and numerical
methods must be used. Note that the addition of a longitudinal field
$B_{z}$ is straightforward, and this calculation can easily be generalized.

\paragraph*{Critical exponents:}

\begin{figure}
\includegraphics[scale=0.8]{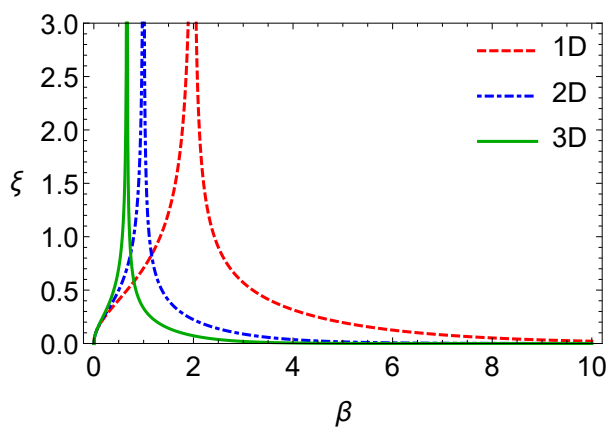}

\caption{\label{fig:Correlation-length}Correlation length $\xi\left(T\right)$
to lowest order in correlations, as a function of the inverse temperature
$\beta$, in units of $V$. At the Curie temperature the correlation
length diverges with critical exponent $\nu=1/2$.}
\end{figure}

Finally, we analyze the effect of correlations on the critical exponents.
As the phase transition is absent in 1D, one must consider $D\geq2$
to discuss the critical exponents. This is a straightforward numerical
calculation which requires to solve the self-consistency equations
Eq.\ref{eq:SCE-CHI-Full} and Eq.\ref{eq:SCE-MZ-Full}. However it
does not provide a clear picture of the effect of correlations, and
for this reason, we will discuss the lowest order solutions only,
which can be done analytically. To lowest order, we can consider Eq.\ref{eq:ZZ-Correlation1}
to describe interspin correlations. In the long wavelength limit of
this expression, the dominant contribution is:
\begin{equation}
\langle S_{k}^{z}S_{-k}^{z}\rangle_{C}\propto\frac{4T_{c}\left(\frac{1}{4}-\bar{M}_{z}^{2}\right)}{4T\cosh^{2}\left(2\bar{M}_{z}\frac{T_{c}}{T}\right)-4T_{c}+Vk^{2}}\label{eq:Corr-B0}
\end{equation}
where $k=\left|\mathbf{k}\right|$ and we have renamed $V_{0}=4T_{c}$.
The Fourier transform of Eq.\ref{eq:Corr-B0} is proportional to $e^{-\frac{\left|x\right|}{\xi}}$,
which corresponds to exponential decay with correlation length:
\begin{equation}
\xi\left(T\right)=\frac{1}{2}\sqrt{\frac{V}{T\cosh^{2}\left(2\bar{M}_{z}\frac{T_{c}}{T}\right)-T_{c}}}
\end{equation}
In Fig.\ref{fig:Correlation-length} we plot the temperature dependence
of the correlation length for the Ising model in different dimensions.
In all cases we find a divergence at the critical temperature $T_{c}$,
and a decrease in $\xi$ as one moves away from the critical point.
To estimate the critical behavior of $\xi$ near $T_{c}$, we expand
in powers of the reduced temperature $t$, and to lowest order we
find:
\begin{equation}
\xi=\frac{1}{2}\sqrt{\frac{V}{T_{c}t}}\rightarrow\nu=1/2
\end{equation}
This is the critical exponent of Ginzburg\textendash Landau theory,
which is expected because we assumed the lowest order correction in
the self-energy, corresponding to first order perturbation (Eq.\ref{eq:ZZ-Correlation1}).
 Similarly, the anomalous dimension $\eta$ can be obtained from the
correlated part of the two-point function at $T=T_{c}$, which according
to scaling arguments\cite{Bellac}, must be proportional to $\langle S^{z}S^{z}\rangle_{k\ll1}^{C}\propto k^{\eta-2}$.
We find from Eq.\ref{eq:Corr-B0}:
\begin{equation}
\langle S_{k}^{z}S_{-k}^{z}\rangle_{C}\propto k^{-2}\rightarrow\eta=0
\end{equation}
meaning that the scaling of fields is not anomalous and agreeing with
the expectations from Ginzburg\textendash Landau theory. If we calculate
the critical exponent $\beta_{t}$ from Eq.\ref{eq:Classical-Corr-Mag},
we find $\beta_{t}=1/2$. This critical exponent has not changed because
we added correlations perturbatively, however the numerical calculation,
including the feedback between the interspin correlations and the
magnetization should display non-perturbative corrections. Nevertheless,
in the next section we will show that even at this level, we can find
some interesting differences between the quantum and the classical
critical point.

\section{Correlations in the Quantum Ising model}

In this section we include the transverse field $B$ and compare with
the previous results. The solutions for the correlation functions
$\mathcal{G}_{\mathbf{k},\mathbf{k}^{\prime}}^{xy\alpha,\beta}$ are
obtained analytically, but their expressions are quite complicated.
Hence we will discuss their general properties and then analyze in
detail two different approximations: the perturbative solution over
the whole phase diagram, and the non-perturbative solution within
the paramagnetic phase. In this last case $B>B_{c}$ and $M_{z}=-i\langle\eta^{x}\eta^{y}\rangle=0$,
highly simplifying the expressions. Importantly, the $T=0$ limit
will allow us to compare the properties of the classical and quantum
critical point.

\begin{figure}
\includegraphics[scale=0.8]{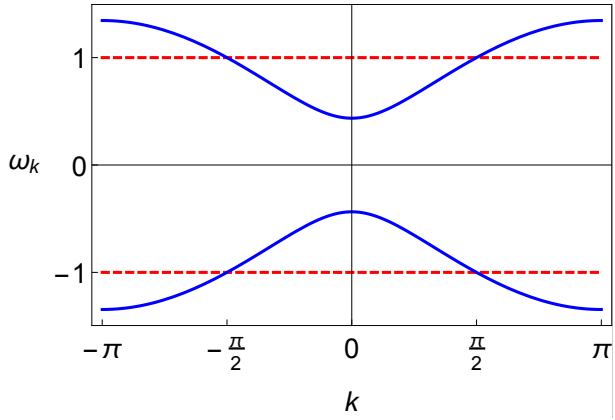}

\caption{\label{fig:Poles}Poles of the Green's function vs $\mathbf{k}$ obtained
from the uncorrelated (dashed, red) and correlated (solid, blue) solution
in 1D. We have chosen $B/V=0.9$ and $\mathcal{Z}=2$ for the plot.
At the critical point the the gap closes.}
\end{figure}

In general, we find that due to the transverse field $B$, the correlation
functions $\mathcal{G}_{\mathbf{k},\mathbf{k}^{\prime}}^{xy\alpha,\beta}$
display new collective excitations called magnons, with dispersion
relationship $\hat{\omega}_{\mathbf{k}}=\sqrt{B^{2}+V_{0}^{2}M_{z}^{2}-BM_{x}V_{\mathbf{k}}}$
(1D case shown in Fig.\ref{fig:Poles}). The dependence on the transverse
magnetization $M_{x}$ makes that, at the critical point $B_{c}=V_{0}/2$,
the gap closes displaying linear dispersion around $\mathbf{k}\simeq0$.
This is related with the scale invariance of the system at low energies,
when $B$ is tuned to the critical value. As in the previous section
we can write the corresponding Dyson's equation:
\begin{equation}
\omega\hat{G}=\frac{\hat{\delta}}{2\pi}+\hat{H}_{0}\cdot\hat{G}+\hat{\Sigma}\cdot\hat{G}+\hat{\Sigma}_{0}\label{eq:Quantum-Full}
\end{equation}
where now $\hat{G}=\left(\begin{array}{ccc}
G_{0}^{x,\beta} & G_{0}^{y,\beta} & G_{0}^{z,\beta}\end{array}\right)^{T}$, $\hat{\delta}=\left(\begin{array}{ccc}
\delta_{x,\beta} & \delta_{y,\beta} & \delta_{z,\beta}\end{array}\right)^{T}$ and
\begin{equation}
\hat{H}_{0}=i\left(\begin{array}{ccc}
0 & M_{z}V_{0} & 0\\
-M_{z}V_{0} & 0 & B\\
0 & -B & 0
\end{array}\right)
\end{equation}
The appearance of two different self-energy contributions $\hat{\Sigma}$
and $\hat{\Sigma}_{0}$ (homogeneous and inhomogeneous term, respectively)
happens because the solutions to $\mathcal{G}_{\mathbf{q},\mathbf{-q}}^{xy\alpha,\beta}$
contain terms proportional to $G_{n,n}^{\alpha,\beta}$ and $G_{nnn,n}^{xyz,\beta}$,
respectively (actually the Green's function $G_{nnn,n}^{xyz,\beta}$
is closely related with the $\Phi$ field defined in ref.\cite{MajoranaRepresentation2}).
Fortunately, $G_{nnn,n}^{xyz,\beta}$ can be calculated exactly from
its equation of motion, and gives:
\begin{eqnarray}
G_{nnn,n}^{xyz,\beta} & = & \frac{i}{2\pi\omega}\left(M_{x}\delta_{x,\beta}+M_{y}\delta_{y,\beta}+M_{z}\delta_{z,\beta}\right)
\end{eqnarray}
Therefore, $\hat{\Sigma}_{0}$ contributes to the source term $\hat{\delta}$,
in addition to the previously discussed change in the quasiparticle
pole coming from $\hat{\Sigma}$. The detailed form of the self-energies
is complicated, and for practical purposes they must be manipulated
numerically.

\begin{figure}
\includegraphics[scale=0.8]{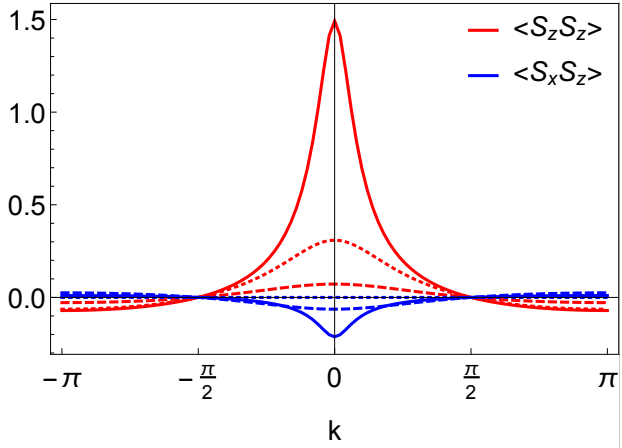}

\caption{\label{fig:Correlations2}Correlated part of the spin-spin correlation
functions $\langle S_{\mathbf{k}}^{\alpha}S_{-\mathbf{k}}^{\beta}\rangle$
vs $\mathbf{k}$ in 1D (lowest order correction). We have chosen $T=0$
and different values for the transverse field $B/V=0.75,\ 0.99\textrm{ and }1.25$
(dashed, solid and dotted, respectively). Correlations increase as
we approach the critical point from both phases, and at the critical
point, the system becomes scale invariant and correlations diverge
at $\mathbf{k}=0$.}
\end{figure}

We first consider the lowest order solution by making the substitution
$\hat{\Sigma}\cdot\hat{G}\rightarrow\hat{\Sigma}\cdot\hat{G}_{0}$
in Eq.\ref{eq:Quantum-Full}. In Fig.\ref{fig:PhaseDiagram1} we plot
the $T=0$ magnetization vs the transverse field. As in the previous
section, the corrections to the magnetization in absence of correlations
are larger in low dimensions, however the corrections are smaller
for the $T=0$ line, and the magnetization in 1D does not drop to
negative values close to the critical point. In Fig.\ref{fig:PhaseDiagram2}
we plot the full phase diagram, where one can see that thermal fluctuations
affect more drastically to the phase diagram, specially as one approaches
the classical critical point $T_{c}$.
Interspin correlations at this order show that, within the ferromagnetic
phase and for $B\neq0$, the crossed correlation function$\langle S_{\mathbf{k}}^{x}S_{\mathbf{k}^{\prime}}^{z}\rangle_{C}\neq0$.
This is what one would expect, as $B$ introduces quantum fluctuations
and $\bar{M}_{x,z}$ are both finite. In Fig.\ref{fig:Correlations2}
we plot the interspin correlation functions in 1D as the transverse
magnetic field increases, crossing the critical point. It shows that
correlations increase as one approaches the phase boundary, diverging
for $\mathbf{k}=0$ at the critical point. We do not show the 2D or
3D case in this plot, as the results are qualitatively the same (we
find the same divergence at $\mathbf{k}=0$ for $B=B_{c}$, and a
slightly different behavior at large $\mathbf{k}$). The difference
in sign between the longitudinal and the transverse correlations happens
because the longitudinal try to align the spins parallel, while the
transverse acts in opposition, trying to break the ferromagnetic order.

Finally, we consider the solution for the paramagnetic phase, by assuming
$B>B_{c}$ and $M_{z}=0$. This allows us to find compact analytical
expressions and gain some insight into the main differences between
the classical and the quantum Ising model. The first advantage of
considering the paramagnetic phase is that the Green's function $G_{0}^{x,\beta}$
decouples from the rest, reducing the matrix dimension. The Dyson's
equation is in this case:
\begin{equation}
\hat{G}=\frac{1}{2\pi}\hat{G}^{\left(0\right)}+\hat{G}^{\left(0\right)}\cdot\hat{\Sigma}\left(\omega\right)\cdot\hat{G}\label{eq:Dyson2}
\end{equation}
where $\hat{G}=\left(\begin{array}{cc}
G_{0}^{y,\beta} & G_{0}^{z,\beta}\end{array}\right)^{T}$, $\hat{\Sigma}\left(\omega\right)=\lambda\left(\omega\right)\left(\begin{array}{cc}
1 & 0\\
0 & 0
\end{array}\right)$ and
\begin{equation}
\lambda\left(\omega\right)=\frac{\omega}{N}\sum_{\mathbf{q}}\left|V_{\mathbf{q}}\right|^{2}\frac{\frac{1}{4}-\frac{1}{N}\langle\eta_{\mathbf{q}}^{x}\eta_{\mathbf{q}}^{y}\eta_{-\mathbf{q}}^{x}\eta_{-\mathbf{q}}^{y}\rangle_{C}}{\omega^{2}-B^{2}+M_{x}BV_{\mathbf{q}}}\label{eq:Self-Energy-PM}
\end{equation}
One important difference between the self-energy in the previous section
(Eq.\ref{eq:Class-Self-energy}) and Eq.\ref{eq:Self-Energy-PM} is
that in this case, the non-commutativity of the interaction and transverse
field terms in the Hamiltonian, mix the integral over $\mathbf{q}$
and the $\omega$ dependence. Then the pole structure for the Green's
function will depend on this integral and in consequence, on the properties
of the interaction $V_{\mathbf{q}}$ and the dimension of the system.
A full solution requires to determine $\langle S_{\mathbf{k}}^{z}S_{\mathbf{k}^{\prime}}^{z}\rangle_{C}$
and solve self-consistently for the magnetization and $\lambda\left(\omega\right)$.
This is why the full solution is significantly more complicated with
the transverse field.

If we consider Eq.\ref{eq:Self-Energy-PM}, one can separate its real
and complex parts $\lambda\left(\omega\pm i\eta\right)=\lambda_{r}\left(\omega\right)\mp i\lambda_{i}\left(\omega\right)$,
being:\begin{widetext}
\begin{eqnarray}
\lambda_{r}\left(\omega\right) & = & \textrm{PV}\frac{\omega}{N}\sum_{\mathbf{q}}\left|V_{\mathbf{q}}\right|^{2}\frac{\frac{1}{4}+\frac{1}{N}\langle S_{\mathbf{q}}^{z}S_{-\mathbf{q}}^{z}\rangle_{C}}{2\sqrt{B^{2}-BM_{x}V_{\mathbf{q}}}}\left(\frac{1}{\omega-\sqrt{B^{2}-BM_{x}V_{\mathbf{q}}}}-\frac{1}{\omega+\sqrt{B^{2}-BM_{x}V_{\mathbf{q}}}}\right)\label{eq:Real-SE}\\
\lambda_{i}\left(\omega\right) & = & \pi\frac{\omega}{N}\sum_{\mathbf{q}}\left|V_{\mathbf{q}}\right|^{2}\frac{\frac{1}{4}+\frac{1}{N}\langle S_{\mathbf{q}}^{z}S_{-\mathbf{q}}^{z}\rangle_{C}}{2\sqrt{B^{2}-BM_{x}V_{\mathbf{q}}}}\left[\delta\left(\omega-\sqrt{B^{2}-BM_{x}V_{\mathbf{q}}}\right)-\delta\left(\omega+\sqrt{B^{2}-BM_{x}V_{\mathbf{q}}}\right)\right]\label{eq:IM-SE}
\end{eqnarray}
\end{widetext}where PV indicates the principal value. Then one can
calculate the formal solution to Eq.\ref{eq:Dyson2} by direct matrix
inversion:
\begin{eqnarray}
G_{0}^{z,y}\left(\omega\right) & = & \frac{-iB}{2\pi\left[\omega^{2}-B^{2}-\omega\lambda\left(\omega\right)\right]}
\end{eqnarray}
which is a function of the self-energy. As the full solution requires
to complement the Green's function with the self-consistency equation
for $M_{x}$, we must obtain its corresponding spectral density:
\begin{eqnarray}
J^{z,y}\left(\omega\right) & = & i\left(e^{\beta\omega}+1\right)^{-1}\left[G_{0}^{z,y}\left(\omega+i\eta\right)-G_{0}^{z,y}\left(\omega-i\eta\right)\right]\nonumber \\
 & = & \frac{1}{\pi}\frac{-i\left(e^{\beta\omega}+1\right)^{-1}B\gamma\left(\omega\right)}{\left[\omega^{2}-B^{2}-m\left(\omega\right)\right]^{2}+\gamma\left(\omega\right)^{2}}\label{eq:SpectralF}
\end{eqnarray}
where we have redefined the the complex and real parts in terms of
the damping $\gamma\left(\omega\right)=\omega\lambda_{i}\left(\omega\right)$
and quasiparticle mass $m\left(\omega\right)=\omega\lambda_{r}\left(\omega\right)$,
respectively. Then the self-consistency equation for $M_{x}$ corresponds
to:
\begin{equation}
iM_{x}=\int_{-\infty}^{\infty}J^{z,y}\left(\omega\right)d\omega\label{eq:MX-SCE}
\end{equation}
To fully characterize the self-energy in Eqs.\ref{eq:Real-SE},\ref{eq:IM-SE}
one needs to obtain an expression for $\langle S_{\mathbf{q}}^{z}S_{-\mathbf{q}}^{z}\rangle_{C}$
from the solution of $\mathcal{G}_{\mathbf{q},-\mathbf{q}}^{xyx,y}$.
An integral form of the solution is not very difficult to obtain,
because the corresponding spectral density can be defined in terms
of $G^{z,y}$ and the contributions from the poles at $\pm\sqrt{B^{2}-BM_{x}V_{\mathbf{q}}}$.
Then Eqs.\ref{eq:Real-SE},\ref{eq:IM-SE} and the self-consistency
equation for $M_{x}$ form a closed set of equations that fully characterize
the paramagnetic phase. The solution needs to be obtained by numerical
means, however one can see from Eq.\ref{eq:SpectralF} that the two
initial quasiparticle poles at $\omega=\pm B$ are now smeared out
by interactions to a Lorentzian shape, and for $\gamma,m\rightarrow0$
one recovers the uncorrelated self-consistency equation. The complex
part of the self-energy $\gamma$ produces damping and the real part
$m$ produces a shift in the quasiparticle energy. If the damping
is small ($\gamma\left(\omega\right)\ll1$) and the spectral function
has maximums at $\omega=E_{\pm}$, with $m\left(\omega\right)$ and
$\gamma\left(\omega\right)$ slow varying functions around them, we
can define the quasiparticles dispersion by solving:
\begin{equation}
E_{\pm}^{2}-B^{2}-m\left(E_{\pm}\right)=0
\end{equation}
This corresponds to approximate quasiparticles (magnons), slightly
damped by interactions and with excitation energies $E_{\pm}$.

To estimate the effect of interactions in the paramagnetic phase,
without solving the full self-consistency equations, we can proceed
in a slightly simpler way if we insert the uncorrelated Green's function
$\hat{G}^{\left(0\right)}$ on the rhs of Eq.\ref{eq:Dyson2} and
solve the self-consistency equation for $T=0$. One finds the next
solution:
\begin{eqnarray}
\langle S_{\mathbf{k}}^{z}S_{\mathbf{k}^{\prime}}^{z}\rangle_{C} & = & -\langle\eta_{\mathbf{k}}^{x}\eta_{\mathbf{k}}^{y}\eta_{\mathbf{k}^{\prime}}^{x}\eta_{\mathbf{k}^{\prime}}^{y}\rangle_{C}\label{eq:Analytical-Corr}\\
 & = & \frac{V_{\mathbf{k}}}{8}\frac{N\delta_{\mathbf{k},-\mathbf{k}^{\prime}}}{B-\frac{V_{\mathbf{k}}}{2}+\sqrt{B^{2}-BM_{x}V_{\mathbf{k}}}}\nonumber 
\end{eqnarray}
Eq.\ref{eq:Analytical-Corr} corresponds to the lowest order solution
for the spin-spin correlation and is valid within the paramagnetic
region of the phase diagram. As expected to lowest order, the divergence
happens at the uncorrelated critical point $B_{c}=V_{0}/2$ (which
is correct in 1D, and therefore should not display the instability
previously found for the RPA, due to the shift of the Curie temperature);
however this divergence shows an important difference with the one
obtained in the classical Ising model (Eq.\ref{eq:ZZ-Correlation1}):
the non-commutativity of the interaction and transverse field terms
in the Hamiltonian produce the square root term in the denominator
of Eq.\ref{eq:Analytical-Corr}, which modifies the scaling properties
near the critical point. To see this explicitly, we consider the long
wavelength limit of Eq.\ref{eq:Analytical-Corr}, as a function of
the reduced field $b=\left(B-B_{c}\right)/B_{c}$:
\begin{equation}
\langle S_{k}^{z}S_{-k}^{z}\rangle_{C}=\frac{1/4}{b+\frac{Vk^{2}}{2B_{c}}+\sqrt{b+1}\sqrt{b+\frac{Vk^{2}}{2B_{c}}}}
\end{equation}
Near $b\sim0$, a change in the behavior of interspin correlations
from $\left(b+\frac{Vk^{2}}{2B_{c}}\right)^{-1}$ to $\left(b+\frac{Vk^{2}}{2B_{c}}\right)^{-1/2}$
will happen for $k^{2}\sim2bB_{c}/V$. The Fourier transform to position
space is in this case proportional to the Bessel function of the second
kind $K_{0}\left(\sqrt{\mathcal{Z}b}\left|r\right|\right)$, which
displays the required asymptotic behavior for interspin correlations
$\sim e^{\sqrt{\mathcal{Z}b}\left|r\right|}$ at large distances.
The difference will be relevant when the distance is of the order
of the inverse correlation length $\left|r\right|\sim\sqrt{b}$ only.
The correlation length near the critical point scales as $\xi\sim b^{-1/2}$,
and we obtain the critical exponent $\nu=1/2$, as in the model without
transverse field; however, the anomalous dimension obtained at the
critical point is:
\begin{align}
\langle S_{k}^{z}S_{-k}^{z}\rangle_{C} & \sim\frac{1}{\sqrt{k^{2}}}\rightarrow\eta=1
\end{align}
This shows that with this method, the perturbative solution over the
uncorrelated solution already captures a difference in the anomalous
dimension between the phase transition driven by quantum fluctuations
and the one driven by thermal fluctuations. Unfortunately we cannot
derive an analytical formula for the scaling of the magnetization
near $B_{c}$, because Eq.\ref{eq:Analytical-Corr} applies to the
PM phase only, where the order parameter vanishes.

Now one can insert Eq.\ref{eq:Analytical-Corr} into Eq.\ref{eq:Self-Energy-PM}.
As previously discussed, this approximation can fail when correlations
are strong and induce an instability of the critical point. However
in this case we know that the $T=0$ critical point in 1D is correct,
and therefore the approximation should be reliable. One obtains:
\begin{equation}
\lambda\left(\omega\right)=\frac{\omega}{2\pi}\int_{-\pi}^{\pi}\frac{\left(\frac{V_{q}}{2}\right)^{2}\left(B+\Omega_{q}\right)dq}{\left(B-\frac{V_{q}}{2}+\Omega_{q}\right)\left(\omega^{2}-\Omega_{q}^{2}\right)}
\end{equation}
where $\Omega_{q}=\sqrt{B^{2}-BM_{x}V_{q}}$ and we have taken the
continuum limit. Adding an infinitesimal complex part $\omega\rightarrow\omega\pm i\epsilon$
the self-energy can then be calculated by contour integral methods.
However, one must notice that the square root dependence introduces
branch cuts which need to be accounted for. Hence, as we are mostly
interested in the frequency dependence, and the condition $B>B_{c}$
removes any divergences coming from $B-\frac{V_{q}}{2}+\Omega_{q}$,
the final expression, to lowest order in $B/V$ for the integrand
$V_{q}^{2}\left(B+\Omega_{q}\right)/\left(B-\frac{V_{q}}{2}+\Omega_{q}\right)$,
should be a good approximation far from the critical point. The self-energy
is then approximated by:
\begin{eqnarray}
\lambda\left(\omega\right) & \simeq & \omega\frac{B^{2}-\omega^{2}}{4B^{2}M_{x}^{2}}-\frac{\omega\left(B^{2}-\omega^{2}\right)^{2}\textrm{sign}\left(B-\left|\omega\right|\right)}{4B^{2}M_{x}^{2}\sqrt{\left(B^{2}-\omega^{2}\right)^{2}-4B^{2}V^{2}M_{x}^{2}}}\label{eq:Approx-SE}
\end{eqnarray}
Eq.\ref{eq:Approx-SE} displays a rich structure as a function of
frequency. The spectral density changes from the two initial poles
$\omega=\pm B$ obtained in absence of correlations, to a combination
of isolated poles and a continuum density of states, once the correlations
are added. Also, at the large field fixed point $B\gg V$, the self-energy
tends to zero and the two poles found in absence of correlations are
recovered, as one would expect. This final result fully characterizes
the solution for $\hat{G}\left(\omega\right)$, and shows how different
approximations can be used on the hierarchy of correlations to obtain
non-perturbative, analytical results.

\section{Conclusions}

We have studied the Ising model using the equation of motion technique
and a decoupling scheme based on the scaling of spatial correlations
between spins. We have shown that this method allows for a general
approach to spin systems, which is not restricted to Ising models
only. Furthermore, a similar approach would apply for the case of
fermionic and bosonic systems as well. For the case of spin systems,
the formalism is specially simple in terms of double-time Green's
functions and Majorana fermions, but it can be applied to different
spin representations such as the Holstein\textendash Primakoff, hardcore
bosons, fermionic, etc, without any difficulties. One of the advantages
is that it allows to discuss systems in different dimensions and with
a large variety of interactions. The only changes will be in the scaling
properties of the hierarchy (its convergence) and the numerical integration
in the self-consistency equations. Importantly, the double-time formalism
can be applied to out-of-time correlators as well\cite{Perk2009},
which expands the possibilities of this method to study dynamical
properties as well.

Regarding the specific models studied in this work, our results show
that the Ising model in absence of a transverse field can be solved
exactly to $\mathcal{Z}^{-1}$ order. This already provides interesting
results, even in 1D where we expect the hierarchy to be less accurate
due to the small coordination number. We demonstrate the absence of
a phase transition in 1D, and predict the value of the spin-spin correlation
function $\langle S_{\mathbf{q}}^{z}S_{-\mathbf{q}}^{z}\rangle_{C}$
and of the energy of the ground state $E_{0}$ (Fig.\ref{fig:SCE-Full}).
Importantly, Eqs.\ref{eq:SCE-MZ-Full},\ref{eq:SCE-CHI-Full} and
\ref{eq:Energy-Class-Ising} correspond to a generalization of the
widely used mean-field equations for the magnetization, now including
two-body correlations. They are valid in arbitrary dimension and allow
to characterize the phase diagram and the critical exponents. In addition,
we have also shown how one can recover the well known Ginzburg-Landau
critical exponents from the perturbative solution to lowest order.

When the transverse field is non-vanishing, the equations of motion
for the Green's functions still can be solved analytically, but their
self-consistency equations are more complicated and require numerical
approaches. In general, one finds the appearance of dispersive collective
modes (magnons) and branch cuts in the spectrum. An interesting difference,
that appears to lowest order in perturbations, is that the critical
exponent $\eta\neq0$ at the quantum critical point. This stresses
the difference with other methods that would just give the mean field
value $\eta=0$. We have shown that interactions in the quantum Ising
model lead, in general, to a complicated expression for the self-energy
including damping, and that the quasiparticle picture becomes just
an approximate one. To show this explicitly we have approximated the
behavior of the self-energy within the paramagnetic phase; it shows
a rich behavior as a function of frequency, where discrete and continuum
excitations are present.

We would like to acknowledge P.C.E. Stamp, T. Cox, and R. McKenzie
for useful discussions. This work was supported by NSER of Canada.

\bibliographystyle{phaip}
\bibliography{hierarchy&Spins}

\begin{widetext}

\appendix

\section{Correlated part of the 4-point function}

Here we include the details for the calculation of the 4-point function
$G_{ppn,n}^{xy\alpha,\beta}\left(t,t^{\prime}\right)=-i\langle\eta_{p}^{x}\left(t\right)\eta_{p}^{y}\left(t\right)\eta_{n}^{\alpha}\left(t\right);\eta_{n}^{\beta}\left(t^{\prime}\right)\rangle$.
The general equation of motion for $p\neq n$ is:
\begin{eqnarray}
\omega G_{ppn,n}^{\mu\nu\alpha,\beta} & = & \frac{1}{2\pi}\langle\left\{ \eta_{p}^{\mu}\eta_{p}^{\nu}\eta_{n}^{\alpha},\eta_{n}^{\beta}\right\} \rangle+i\langle\left[H,\eta_{p}^{\mu}\eta_{p}^{\nu}\eta_{n}^{\alpha}\right];\eta_{n}^{\beta}\rangle
\end{eqnarray}
As we are interested in the correlated part of the Green's function
we must remove the equation of motion for the uncorrelated part $\mathcal{G}_{ppn,n}^{\mu\nu\alpha,\beta}=G_{ppn,n}^{\mu\nu\alpha,\beta}-\langle\eta_{p}^{\mu}\eta_{p}^{\nu}\rangle G_{n,n}^{\alpha,\beta}$:
\begin{eqnarray}
\omega\mathcal{G}_{ppn,n}^{\mu\nu\alpha,\beta} & = & \epsilon_{x\alpha\theta}iB\left(G_{ppn,n}^{\mu\nu\theta,\beta}-\langle\eta_{p}^{\mu}\eta_{p}^{\nu}\rangle G_{n,n}^{\theta,\beta}\right)+iB\left(\epsilon_{x\mu\theta}G_{ppn,n}^{\theta\nu\alpha,\beta}+\epsilon_{x\nu\theta}G_{ppn,n}^{\mu\theta\alpha,\beta}\right)\\
 &  & +\epsilon_{z\alpha\theta}\sum_{i\neq n}V_{n,i}\left(G_{ppiin,n}^{\mu\nu xy\theta,\beta}-\langle\eta_{p}^{\mu}\eta_{p}^{\nu}\rangle G_{iin,n}^{xy\theta,\beta}\right)+\sum_{i\neq p}V_{p,i}\left(\epsilon_{z\mu\theta}G_{iippn,n}^{xy\theta\nu\alpha,\beta}+\epsilon_{z\nu\theta}G_{iippn,n}^{xy\mu\theta\alpha,\beta}\right)\nonumber 
\end{eqnarray}
Then, we make use of the conservation law $\partial_{t}\langle\eta_{p}^{\mu}\eta_{p}^{\nu}\rangle=0$
and expand in correlations (neglecting $\mathcal{Z}^{-2}$ order terms),
finding:
\begin{eqnarray}
\omega\mathcal{G}_{ppn,n}^{\mu\nu\alpha,\beta} & = & iB\left(\epsilon_{x\alpha\theta}\mathcal{G}_{ppn,n}^{\mu\nu\theta,\beta}+\epsilon_{x\mu\theta}\mathcal{G}_{ppn,n}^{\theta\nu\alpha,\beta}+\epsilon_{x\nu\theta}\mathcal{G}_{ppn,n}^{\mu\theta\alpha,\beta}\right)\\
 &  & +\epsilon_{z\alpha\theta}\sum_{i\neq n,p}V_{n,i}\left(\langle\eta_{p}^{\mu}\eta_{p}^{\nu}\eta_{i}^{x}\eta_{i}^{y}\rangle_{C}G_{n,n}^{\theta,\beta}+\langle\eta_{i}^{x}\eta_{i}^{y}\rangle\mathcal{G}_{ppn,n}^{\mu\nu\theta,\beta}\right)\nonumber \\
 &  & +\epsilon_{z\alpha\theta}V_{n,p}\left(\langle\eta_{p}^{\mu}\eta_{p}^{\nu}\eta_{p}^{x}\eta_{p}^{y}\rangle-\langle\eta_{p}^{\mu}\eta_{p}^{\nu}\rangle\langle\eta_{p}^{x}\eta_{p}^{y}\rangle\right)G_{n,n}^{\theta,\beta}\nonumber \\
 &  & +\epsilon_{z\mu\theta}\sum_{i\neq p,n}V_{p,i}\left(\langle\eta_{i}^{x}\eta_{i}^{y}\rangle\mathcal{G}_{ppn,n}^{\theta\nu\alpha,\beta}+\langle\eta_{p}^{\theta}\eta_{p}^{\nu}\rangle\mathcal{G}_{iin,n}^{xy\alpha,\beta}\right)\nonumber \\
 &  & +\epsilon_{z\nu\theta}\sum_{i\neq p,n}V_{p,i}\left(\langle\eta_{i}^{x}\eta_{i}^{y}\rangle\mathcal{G}_{ppn,n}^{\mu\theta\alpha,\beta}+\langle\eta_{p}^{\mu}\eta_{p}^{\theta}\rangle\mathcal{G}_{iin,n}^{xy\alpha,\beta}\right)\nonumber \\
 &  & +V_{p,n}\left(\epsilon_{z\mu\theta}\langle\eta_{p}^{\theta}\eta_{p}^{\nu}\rangle+\epsilon_{z\nu\theta}\langle\eta_{p}^{\mu}\eta_{p}^{\theta}\rangle\right)\left(G_{nnn,n}^{xy\alpha,\beta}-\langle\eta_{n}^{x}\eta_{n}^{y}\rangle G_{n,n}^{\alpha,\beta}\right)\nonumber 
\end{eqnarray}
Importantly, if we make use of the properties of the transverse Ising
model $\langle S_{i}^{y}\rangle=0$ and $\langle S_{i}^{z}S_{j}^{y}\rangle=0$,
we can simplify the different concrete cases. Also, the Green's function
$G_{nnn,n}^{xy\alpha,\beta}$ appear in the equation of motion, but
fortunately it can be calculated exactly:
\begin{align}
G_{nnn,n}^{xyz,\beta} & =\frac{1}{2\pi\omega}\left(\delta_{x,\beta}\langle\eta_{n}^{y}\eta_{n}^{z}\rangle-\delta_{y,\beta}\langle\eta_{n}^{x}\eta_{n}^{z}\rangle+\delta_{z,\beta}\langle\eta_{n}^{x}\eta_{n}^{y}\rangle\right)
\end{align}

\section{Classical Ising model}

The full equation of motion in momentum space is given by:
\begin{eqnarray}
\omega G_{\mathbf{k}}^{\alpha,\beta} & = & \frac{N}{2\pi}\delta\left(\mathbf{k}\right)\delta_{\alpha,\beta}+iB\epsilon_{x\alpha\mu}G_{\mathbf{k}}^{\mu,\beta}+\epsilon_{z\alpha\mu}iM_{z}V_{0}G_{\mathbf{k}}^{\mu,\beta}+\epsilon_{z\alpha\mu}\frac{1}{N}\sum_{\mathbf{q}}V_{\mathbf{q}}\mathcal{G}_{\mathbf{q},\mathbf{k}-\mathbf{q}}^{xy\mu,\beta}
\end{eqnarray}
Neglecting the transverse term and writing the different components
in matrix form we find:
\begin{equation}
\left(\begin{array}{cc}
\omega & -iM_{z}V_{0}\\
iM_{z}V_{0} & \omega
\end{array}\right)\left(\begin{array}{c}
G_{\mathbf{k}}^{x,\beta}\\
G_{\mathbf{k}}^{y,\beta}
\end{array}\right)=N\frac{\delta_{\mathbf{k},0}}{2\pi}\left(\begin{array}{c}
\delta_{x,\beta}\\
\delta_{y,\beta}
\end{array}\right)+\frac{1}{N}\sum_{\mathbf{q}}V_{\mathbf{q}}\left(\begin{array}{c}
\mathcal{G}_{\mathbf{q},\mathbf{k}-\mathbf{q}}^{xyy,\beta}\\
-\mathcal{G}_{\mathbf{q},\mathbf{k}-\mathbf{q}}^{xyx,\beta}
\end{array}\right)
\end{equation}
Solving the equation of motion for the correlated part and assuming
that the magnetization is homogeneous we find:
\begin{align*}
\mathcal{G}_{\mathbf{k},\mathbf{k}^{\prime}}^{xyx,\beta} & =V_{\mathbf{k}^{\prime}}\frac{N\delta_{\mathbf{k},-\mathbf{k}^{\prime}}\left(\frac{1}{4}-M_{z}^{2}\right)-\langle\eta_{\mathbf{k}}^{x}\eta_{\mathbf{k}}^{y}\eta_{\mathbf{k}^{\prime}}^{x}\eta_{\mathbf{k}^{\prime}}^{y}\rangle_{C}}{\omega^{2}-V_{0}^{2}M_{z}^{2}}\left(iV_{0}M_{z}G_{0}^{x,\beta}-\omega G_{0}^{y,\beta}\right)\\
\mathcal{G}_{\mathbf{k},\mathbf{k}^{\prime}}^{xyy,\beta} & =V_{\mathbf{k}^{\prime}}\frac{N\delta_{\mathbf{k},-\mathbf{k}^{\prime}}\left(\frac{1}{4}-M_{z}^{2}\right)-\langle\eta_{\mathbf{k}}^{x}\eta_{\mathbf{k}}^{y}\eta_{\mathbf{k}^{\prime}}^{x}\eta_{\mathbf{k}^{\prime}}^{y}\rangle_{C}}{\omega^{2}-V_{0}^{2}M_{z}^{2}}\left(iV_{0}M_{z}G_{0}^{y,\beta}+\omega G_{0}^{x,\beta}\right)
\end{align*}
where $G_{\mathbf{k}}^{\alpha,\beta}=G_{0}^{\alpha,\beta}\delta\left(\mathbf{k}\right)/N$.
Then we can rewrite the initial equation of motion as:
\begin{equation}
\left(\omega-\hat{H}_{0}\right)\hat{G}=\frac{\hat{\delta}}{2\pi}+\hat{\Sigma}\left(\omega\right)\cdot\hat{G}
\end{equation}
being $\hat{G}=\left(\begin{array}{c}
G_{0}^{x,\beta}\\
G_{0}^{y,\beta}
\end{array}\right)$, $\hat{\delta}=\left(\begin{array}{c}
\delta_{x,\beta}\\
\delta_{y,\beta}
\end{array}\right)$, $\hat{H}_{0}=\left(\begin{array}{cc}
0 & iM_{z}V_{0}\\
-iM_{z}V_{0} & 0
\end{array}\right)$, $\hat{G}^{\left(0\right)}=\left(\omega-\hat{H}_{0}\right)^{-1}\cdot\hat{\delta}$
and
\begin{eqnarray}
\hat{\Sigma}\left(\omega\right) & = & \lambda\left(\omega\right)\left(\begin{array}{cc}
\omega & iV_{0}M_{z}\\
-iV_{0}M_{z} & \omega
\end{array}\right)\\
\lambda\left(\omega\right) & = & \frac{1}{N}\sum_{\mathbf{q}}\left|V_{\mathbf{q}}\right|^{2}\frac{\frac{1}{4}-M_{z}^{2}-\frac{1}{N}\langle\eta_{\mathbf{q}}^{x}\eta_{\mathbf{q}}^{y}\eta_{-\mathbf{q}}^{x}\eta_{-\mathbf{q}}^{y}\rangle_{C}}{\omega^{2}-V_{0}^{2}M_{z}^{2}}\nonumber \\
 & = & \frac{\chi}{\omega^{2}-V_{0}^{2}M_{z}^{2}}
\end{eqnarray}
where $\chi=\frac{1}{N}\sum_{\mathbf{q}}\left|V_{\mathbf{q}}\right|^{2}\left[\frac{1}{4}-M_{z}^{2}-\frac{1}{N}\langle\eta_{\mathbf{q}}^{x}\eta_{\mathbf{q}}^{y}\eta_{-\mathbf{q}}^{x}\eta_{-\mathbf{q}}^{y}\rangle_{C}\right]$.
Note that this previous expression indicates that a full solution
requires to calculate $\chi$ later on, in a self-consistent way.
The solution for the Green's function is easily found to be:
\begin{equation}
\hat{G}=\left(\omega-\hat{H}_{0}-\hat{\Sigma}\left(\omega\right)\right)^{-1}\cdot\frac{\hat{\delta}}{2\pi}
\end{equation}
with poles at:
\begin{equation}
\omega_{i}=\pm M_{z}V_{0}\pm\sqrt{\chi}
\end{equation}
Then we calculate the self-consistency equations for the magnetization
and the interspin correlations. For the magnetization we find:
\begin{equation}
M_{z}=\frac{1}{2}\frac{\sinh\left(\frac{M_{z}V_{0}}{T}\right)}{\cosh\left(\frac{M_{z}V_{0}}{T}\right)+\cosh\left(\frac{\sqrt{\chi}}{T}\right)}
\end{equation}
To correctly determine $\chi$ we need to first find a self-consistency
equation for $\langle\eta_{\mathbf{q}}^{x}\eta_{\mathbf{q}}^{y}\eta_{-\mathbf{q}}^{x}\eta_{-\mathbf{q}}^{y}\rangle_{C}$,
which is obtained from the expression for $\mathcal{G}_{\mathbf{q},-\mathbf{q}}^{xyx,y}$.
We find:
\[
-\langle\eta_{\mathbf{q}}^{x}\eta_{\mathbf{q}}^{y}\eta_{-\mathbf{q}}^{x}\eta_{-\mathbf{q}}^{y}\rangle_{C}=V_{-\mathbf{q}}\frac{N\left(\frac{1}{4}-M_{z}^{2}\right)-\langle\eta_{\mathbf{q}}^{x}\eta_{\mathbf{q}}^{y}\eta_{-\mathbf{q}}^{x}\eta_{-\mathbf{q}}^{y}\rangle_{C}}{2\sqrt{\chi}\left[\cosh\left(\frac{M_{z}V_{0}}{T}\right)+\cosh\left(\frac{\sqrt{\chi}}{T}\right)\right]}\sinh\left(\frac{\sqrt{\chi}}{T}\right)
\]
which corresponds to :
\begin{equation}
\langle\eta_{\mathbf{q}}^{x}\eta_{\mathbf{q}}^{y}\eta_{-\mathbf{q}}^{x}\eta_{-\mathbf{q}}^{y}\rangle_{C}=-\frac{N\left(\frac{1}{4}-M_{z}^{2}\right)V_{-\mathbf{q}}\sinh\left(\frac{\sqrt{\chi}}{T}\right)}{2\sqrt{\chi}\left[\cosh\left(\frac{M_{z}V_{0}}{T}\right)+\cosh\left(\frac{\sqrt{\chi}}{T}\right)\right]-V_{-\mathbf{q}}\sinh\left(\frac{\sqrt{\chi}}{T}\right)}
\end{equation}
Note that it nicely agrees with the lowest order, perturbative solution
$\langle\eta_{\mathbf{q}}^{x}\eta_{\mathbf{q}}^{y}\eta_{-\mathbf{q}}^{x}\eta_{-\mathbf{q}}^{y}\rangle_{c}=\frac{-N\left(\frac{1}{4}-M_{z}^{2}\right)V_{-q}}{4T\cosh^{2}\left(\frac{V_{0}M_{z}}{2T}\right)-V_{-q}}$
if we expand to linear order in $\chi$. To finally obtain the self-consistency
equation for $\chi$, we insert the previous result into $\chi=\frac{1}{N}\sum_{\mathbf{q}}\left|V_{\mathbf{q}}\right|^{2}\left[\frac{1}{4}-M_{z}^{2}-\frac{1}{N}\langle\eta_{\mathbf{q}}^{x}\eta_{\mathbf{q}}^{y}\eta_{-\mathbf{q}}^{x}\eta_{-\mathbf{q}}^{y}\rangle_{C}\right]$,
and get:
\begin{eqnarray}
\chi & = & \left(\frac{1}{4}-M_{z}^{2}\right)\frac{1}{N}\sum_{\mathbf{q}}\frac{2\sqrt{\chi}\left[\cosh\left(\frac{M_{z}V_{0}}{T}\right)+\cosh\left(\frac{\sqrt{\chi}}{T}\right)\right]\left|V_{\mathbf{q}}\right|^{2}}{2\sqrt{\chi}\left[\cosh\left(\frac{M_{z}V_{0}}{T}\right)+\cosh\left(\frac{\sqrt{\chi}}{T}\right)\right]-V_{-\mathbf{q}}\sinh\left(\frac{\sqrt{\chi}}{T}\right)}
\end{eqnarray}
which correctly agrees with the RPA result to linear order in $\chi$.
In the 1D case one can obtain an exact expression for the integral:
\[
\int_{-\pi}^{\pi}\frac{dq}{2\pi}\frac{\cos^{2}\left(q\right)}{m-\cos\left(q\right)}=\frac{m\theta\left(m-1\right)}{m^{2}-1+m\sqrt{m^{2}-1}}-m\theta\left(1-m\right)
\]
where we have computed its value for $m>1$ by contour integral in
the complex plane, and to extend the result to $m<1$, we have ``regularized''
the integral by calculating its principal value, deforming the unit
circle around the poles. The final result is
\begin{eqnarray}
\chi & \overset{1D}{=} & \left(\frac{1}{4}-M_{z}^{2}\right)V_{0}^{2}m^{2}\left(\frac{\theta\left(m-1\right)}{m^{2}-1+m\sqrt{m^{2}-1}}-\theta\left(1-m\right)\right)
\end{eqnarray}
being $m=2\sqrt{\chi}\frac{\cosh\left(\frac{M_{z}V_{0}}{T}\right)+\cosh\left(\frac{\sqrt{\chi}}{T}\right)}{V_{0}\sinh\left(\frac{\sqrt{\chi}}{T}\right)}$.
The last important result is the calculation of the average energy,
which in case of different possible solutions, will determine the
ground state. The energy per spin of the ground state is:
\begin{equation}
\frac{E_{0}}{N}=\frac{\langle H\rangle}{N}=-\frac{1}{N}\sum_{i,j>i}V_{i,j}\langle S_{i}^{z}S_{j}^{z}\rangle
\end{equation}
In the Majorana representation, the energy per spin becomes:
\begin{eqnarray}
\frac{E_{0}}{N} & = & \frac{1}{N}\sum_{i,j>i}V_{i,j}\langle\eta_{i}^{x}\eta_{i}^{y}\eta_{j}^{x}\eta_{j}^{y}\rangle\\
 & = & V_{0}\langle\eta^{x}\eta^{y}\rangle^{2}+\frac{1}{N}\sum_{\mathbf{q}}V_{\mathbf{q}}\langle\eta_{\mathbf{q}}^{x}\eta_{\mathbf{q}}^{y}\eta_{-\mathbf{q}}^{x}\eta_{-\mathbf{q}}^{y}\rangle_{c}
\end{eqnarray}
where in the second line we have separated into uncorrelated and correlated
parts, and assumed homogeneous magnetization. The magnetization is
obtained from the solution of the self-consistency equation, and the
last term can be related with the two-point function as: 
\begin{eqnarray}
\frac{E_{0}}{N} & = & V_{0}\langle\eta^{x}\eta^{y}\rangle^{2}+i\int_{-\infty}^{\infty}\frac{\omega G_{0}^{y,y}\left(\omega\pm i\eta\right)+iV_{0}M_{z}G_{0}^{x,y}\left(\omega\pm i\eta\right)}{e^{\beta\omega}+1}d\omega
\end{eqnarray}
resulting in
\begin{eqnarray}
\frac{E_{0}}{N} & = & -V_{0}M_{z}^{2}-\frac{1}{2}\frac{\sqrt{\chi}\sinh\left(\frac{\sqrt{\chi}}{T}\right)}{\cosh\left(\frac{M_{z}V_{0}}{T}\right)+\cosh\left(\frac{\sqrt{\chi}}{T}\right)}
\end{eqnarray}

\end{widetext}
\end{document}